\documentstyle[12pt,epsf]{article}
\def\fo{\hbox{{1}\kern-.25em\hbox{l}}}
\def\fnote#1#2{\begingroup\def\thefootnote{#1}\footnote{#2}\addtocounter
{footnote}{-1}\endgroup}

\renewcommand{\thefootnote}{\fnsymbol{footnote}}

\def\beq{\begin{equation}}
\def\eeq{\end{equation}}
\def\eq{\end{equation}}
\def\to{\rightarrow}

\def\bsg{\ifmmode B\to X_s\gamma\else $B\to X_s\gamma$\fi}
\def\bsll{\ifmmode B\to X_s\ell^+\ell^-\else $B\to X_s\ell^+\ell^-$\fi}
\def\bstt{\ifmmode B\to X_s\tau^+\tau^-\else $B\to X_s\tau^+\tau^-$\fi}
\def\shat{\ifmmode \hat{s}\else $\hat{s}$\fi}

\newcommand{\newc}{\newcommand}

\newc{\lcal}{\int {\cal L}dt}
 
\newc{\mstop}{m_{\tilde{t}}}
\newc{\mHpm}{m_{H^\pm}}
\newc{\gsim}{\lower.7ex\hbox{$\;\stackrel{\textstyle>}{\sim}\;$}}
\newc{\lsim}{\lower.7ex\hbox{$\;\stackrel{\textstyle<}{\sim}\;$}}
\newc{\ie}{{\it i.e.}}          
\newc{\etal}{{\it et al.}}
\newc{\eg}{{\it e.g.}}          
\newc{\kev}{\hbox{\rm\,keV}}            
\newc{\mev}{\hbox{\rm\,MeV}}            
\newc{\gev}{\hbox{\rm\,GeV}}            
\newc{\tev}{\hbox{\rm\,TeV}}
\newc{\xpb}{\hbox{\rm\, pb}}
\newc{\xfb}{\hbox{\rm\, fb}}

\def\alt{\stackrel{<}{\sim}}

\newc{\mtop}{m_t}
\newc{\mbot}{m_b}
\newc{\mz}{m_Z}
\newc{\mw}{M_W}
\newc{\alphasmz}{\alpha_s(m_Z^2)}
\newc{\swsq}{\sin^2\theta_W}
\newc{\tw}{\tan\theta_W}
\newc{\cw}{\cos\theta_W}
\newc{\sw}{\sin\theta_W}
\newc{\BR}{\hbox{\rm BR}}
\newc{\zbb}{Z\to b\bar}
\newc{\Gb}{\Gamma (Z\to b\bar b)}
\newc{\Gh}{\Gamma (Z\to \hbox{\rm hadrons})}
\newc{\rbsm}{R_b^\hbox{\rm sm}}
\newc{\rbsusy}{R_b^\hbox{\rm susy}}
\newc{\drb}{\delta R_b}

\newc{\sgn}{\mbox{sgn}}
\newc{\tbeta}{\tan\beta}
\newc{\uL}{{\tilde u_L}}
\newc{\uR}{{\tilde u_R}}
\newc{\cL}{{\tilde c_L}}
\newc{\cR}{{\tilde c_R}}
\newc{\tL}{{\tilde t_L}}
\newc{\tR}{{\tilde t_R}}
\newc{\dL}{{\tilde d_L}}
\newc{\dR}{{\tilde d_R}}
\newc{\sL}{{\tilde s_L}}
\newc{\sR}{{\tilde s_R}}
\newc{\bL}{{\tilde b_L}}
\newc{\bR}{{\tilde b_R}}
\newc{\eL}{{\tilde e_L}}
\newc{\eR}{{\tilde e_R}}
\newc{\mhp}{m_{H^\pm}}
\newc{\mhalf}{m_{1/2}}

\newc{\lR}{\tilde{l}_R}
\newc{\lL}{\tilde{l}_L}
\newc{\nL}{\tilde{\nu}_L}
\newc{\na}{\chi^0_1}
\newc{\nb}{\chi^0_2}
\newc{\nc}{\chi^0_3}
\newc{\nd}{\chi^0_4}
\newc{\ca}{\chi^{\pm}_1}
\newc{\cb}{\chi^{\pm}_2}
\newc{\camp}{\chi^\mp_1}
\newc{\cbmp}{\chi^\mp_1}
\newc{\capos}{\chi^{+}_1}
\newc{\caneg}{\chi^{-}_1}
\newc{\phit}{\phi_t}
\newc{\phib}{\phi_b}
\newc{\phiew}{\phi_{ew}}
\newc{\htz}{h^0_t}
\newc{\hbz}{h^0_b}
\newc{\hewz}{h^0_{ew}}
\newc{\hsmz}{h^0_{sm}}
\newc{\huz}{h^0_u}
\newc{\hsusyz}{h^0_{susy}}

\def\NPB#1#2#3{Nucl. Phys. B {\bf #1}, #3 (19#2)}
\def\PLB#1#2#3{Phys. Lett. B {\bf #1}, #3 (19#2)}

\def\PRD#1#2#3{Phys. Rev. D {\bf #1}, #3 (19#2)}
\def\PRL#1#2#3{Phys. Rev. Lett. {\bf#1}, #3 (19#2)}

\def\ZPC#1#2#3{Zeit. f\"ur Physik C {\bf #1}, #3 (19#2)}

\def\beq{\begin{equation}}
\def\eeq{\end{equation}}
\def\bea{\begin{eqnarray}}
\def\eea{\end{eqnarray}}
\def\slashchar#1{\setbox0=\hbox{$#1$}           
   \dimen0=\wd0                                 
   \setbox1=\hbox{/} \dimen1=\wd1               
   \ifdim\dimen0>\dimen1                        
      \rlap{\hbox to \dimen0{\hfil/\hfil}}      
      #1                                        
   \else                                        
      \rlap{\hbox to \dimen1{\hfil$#1$\hfil}}   
      /                                         
   \fi}                                         %
\catcode`@=11
\long\def\@caption#1[#2]#3{\par\addcontentsline{\csname
  ext@#1\endcsname}{#1}{\protect\numberline{\csname
  the#1\endcsname}{\ignorespaces #2}}\begingroup
    \small
    \@parboxrestore
    \@makecaption{\csname fnum@#1\endcsname}{\ignorespaces #3}\par
  \endgroup}
\catcode`@=12

\def\jfig#1#2#3{
 \begin{figure}
 \centering
 \epsfysize=3.0in
 \hspace*{0in}
 \epsffile{#2}
 \caption{#3}
 \label{#1}
 \end{figure}}

\begin{document}

\begin{titlepage}

\begin{flushleft}
\end{flushleft}
\begin{flushright}
SLAC-PUB-7678 \\
FSU-HEP-971017 \\
hep-ph/9710368\\
October 1997
\end{flushright}
\bigskip



\huge
\begin{center}
Trilepton Higgs signal at hadron colliders
\end{center}

\large

\vspace{.15in}
\begin{center}

Howard Baer${}^{a\dagger}$ and James D.~Wells${}^b$\fnote{\dagger}{Work 
supported by the Department of Energy
under contract DE-AC03-76SF00515 and DE-FG-05-87ER40319.} \\

\vspace{.1in}
{\it ${}^{(a)}$Department of Physics, Florida State University \\
Tallahassee, Florida 32306 \\}
\bigskip
{\it ${}^{(b)}$Stanford Linear Accelerator Center, Stanford University \\
Stanford, California 94309 \\}

\end{center}
 
 
\vspace{0.15in}
 
\begin{abstract}

Determining the origin of electroweak symmetry breaking will be one of
the primary functions of high energy colliders.  We point out that most
Higgs boson searches pursued at hadron colliders require Yukawa
interactions either in the production or the decay of a Higgs boson.
We propose a trilepton Higgs boson search based only upon the gauge
interactions of the Higgs.  This strategy can be utilized successfully for the
standard model (SM) Higgs boson as well as non-standard Higgs bosons 
which break
electroweak symmetry but have little to do with fermion mass
generation.  The trileptons come from $Wh$ production followed
by $Wh\to WWW^{(*)}\to 3l$ decays. A SM Higgs trilepton signal 
would be difficult to
detect at the Tevatron collider: with $100\xfb^{-1}$ of data,
only a $3\sigma$ signal above background can be achieved after cuts
if $140\gev < m_{\hsmz} < 175\gev$.
Some discrimination of signal over background can be gained
by analyzing the opposite sign dilepton $p_T$ distributions.
At the LHC with 30 (100)$\xfb^{-1}$ a clear discovery
above the $5\sigma$ level is possible for a Higgs mass in the range 
$140-185$ ($125-200$)\gev.  Prospects for a trilepton Higgs discovery
are greatly improved for models with non-standard Higgs sectors 
where a Higgs boson couples preferentially
to vector bosons rather than to fermions.

\end{abstract}

\end{titlepage}

\baselineskip=18pt




\vfill
\eject

\section{Introduction}
\bigskip

The mechanism for electroweak symmetry breaking is still a mystery.
The standard model solution of a single condensing Higgs boson
doublet is merely a postulate which presently
happens to not be in contradiction with data.  
The standard model Higgs ($\hsmz$) signatures at lepton, photon and hadron
colliders have been thoroughly studied\cite{review,kunszt}.  Important 
modes of discovery have been identified, and it appears likely that
a standard model Higgs boson will be seen up to about $1\tev$ at
the CERN LHC.

Discovery of a light $\hsmz$ is best accomplished at the CERN LEP2
$e^+e^-$ collider, which ought ultimately to be sensitive to
$m_{\hsmz} \alt M_Z$.
If $160\ {\rm GeV}\alt m_{\hsmz} \alt 800$ GeV, 
then discovery should be possible at the CERN LHC
by searching for the ``gold-plated'' decay $\hsmz\to ZZ^{(*)}\to 4l$
which allows for a Higgs boson mass reconstruction.
If $M_Z\alt m_{\hsmz} \alt 160$ GeV (the case of an intermediate 
mass Higgs boson), then
discovery is perhaps best accomplished at the CERN LHC via a search
for $\hsmz\rightarrow\gamma\gamma$\cite{wudka}.
In the process $gg\to \hsmz\to \gamma\gamma$ the total cross section
peaks at about $50\xfb$ for $m_{\hsmz}\simeq 130\gev$.  The cross-section
reduces to about $25\xfb$ at $m_{\hsmz}\simeq 150\gev$.  The background
is continuum $q\bar q\to \gamma\gamma$ production, which can be overcome
with excellent photon energy resolution -- a feature planned for both
detectors at the LHC~\cite{lhctdr}. 

A SM Higgs boson discovery could also be possible at the Fermilab Tevatron
$p\bar p$ collider. Direct $s$-channel Higgs boson production 
at the Tevatron does not lead to any signals observable above background. 
However, by focusing on Higgs boson 
production in association with a vector boson
($q\bar Q\rightarrow W\hsmz$), a mass bump from $\hsmz\rightarrow b\bar b$
can be reconstructed above background by also tagging on a charged lepton from
the $W$ decay.
With $30\xfb^{-1}$ of integrated luminosity it appears possible to
discover the Higgs in the $l\bar bb$ mode at the Tevatron
if its mass is below
about $120\gev$~\cite{stange,kim}!
Discovery of a SM Higgs via the associated production mechanism is also
possible at the CERN LHC~\cite{agrawal} for 
$m_{\hsmz}\lsim 125\gev$ if an integrated luminosity of
$30\xfb^{-1}$ is accumulated. 

Other topology searches can discover and effectively
measure the mass of the light standard model Higgs boson.  However, 
one should carefully study all possible 
modes of detecting the degrees of freedom
arising from electroweak symmetry breaking since we do not presently
know exactly how this symmetry breaking is accomplished or how it will show
up experimentally.  In some cases,
the secondary or tertiary modes for detecting the standard model Higgs
boson become the most important discovery modes for the actual mechanism 
nature has chosen. Therefore we study another method to search for
the Higgs boson. Namely, as the intermediate mass Higgs boson gets heavier,
its decays into $WW^{(*)}$ get larger and can become relevant.  
Sometimes these decays will lead to two leptons and missing energy
from $\hsmz\to WW^{(*)}\to l\nu l\nu$.  If the associated $W$ decays
leptonically one is left with a trilepton signature with missing
energy: $W\hsmz\to WWW^{(*)}\to 3l+\slashchar{E_T}$.  The backgrounds
to this process are small but important and will be discussed in the sections
on the Tevatron and LHC search capabilities below.  

The trilepton Higgs mode is unique among Higgs signatures at
hadron colliders in that it occurs only by gauge interactions.
Some studies have been carried out for 
$Wh\to l\nu\gamma\gamma$~\cite{stirling,stange,akeroyd},
which is also allowed by purely gauge interactions.  However,
this process is usually only applicable for Higgs masses below
$100\gev$ -- a mass region that can be covered effectively by
LEP2 searches in the next year.
The $gg\to \hsmz \to \gamma\gamma$ requires Yukawa interactions
for the production, $pp\to W^\pm \hsmz\to l\nu b\bar b$ requires
Yukawa interactions in the decay, and even $gg\to \hsmz\to WW^*$
studied in~\cite{glover,barger90,dittmar}, which utilizes the
same $\hsmz\to WW^*\to l\nu l\nu$ decay, requires Yukawa
interactions in the production (unless $m_{\hsmz}$ is very
high and then $WW$ fusion becomes important).  Since
$pp\to W^\pm \hsmz\to 3l$ only requires gauge interactions it
could probe some Higgs bosons that the other detection modes could not if
there are scalar degrees of freedom associated with electroweak
symmetry breaking that have little to do with fermion mass
generation.  It is partly for this reason why we consider it
important to study this trilepton signal.

\section{Higgs decays to leptons}
\bigskip

In the standard model the branching fractions of the Higgs boson are
presented in Fig.~\ref{br-hsm}.  The $b\bar b$ mode dominates up
to about $130\gev$.  Above $130\gev$ the $WW^{(*)}$ becomes competitive
and eventually dominates the decay modes of the standard model
Higgs.
\jfig{br-hsm}{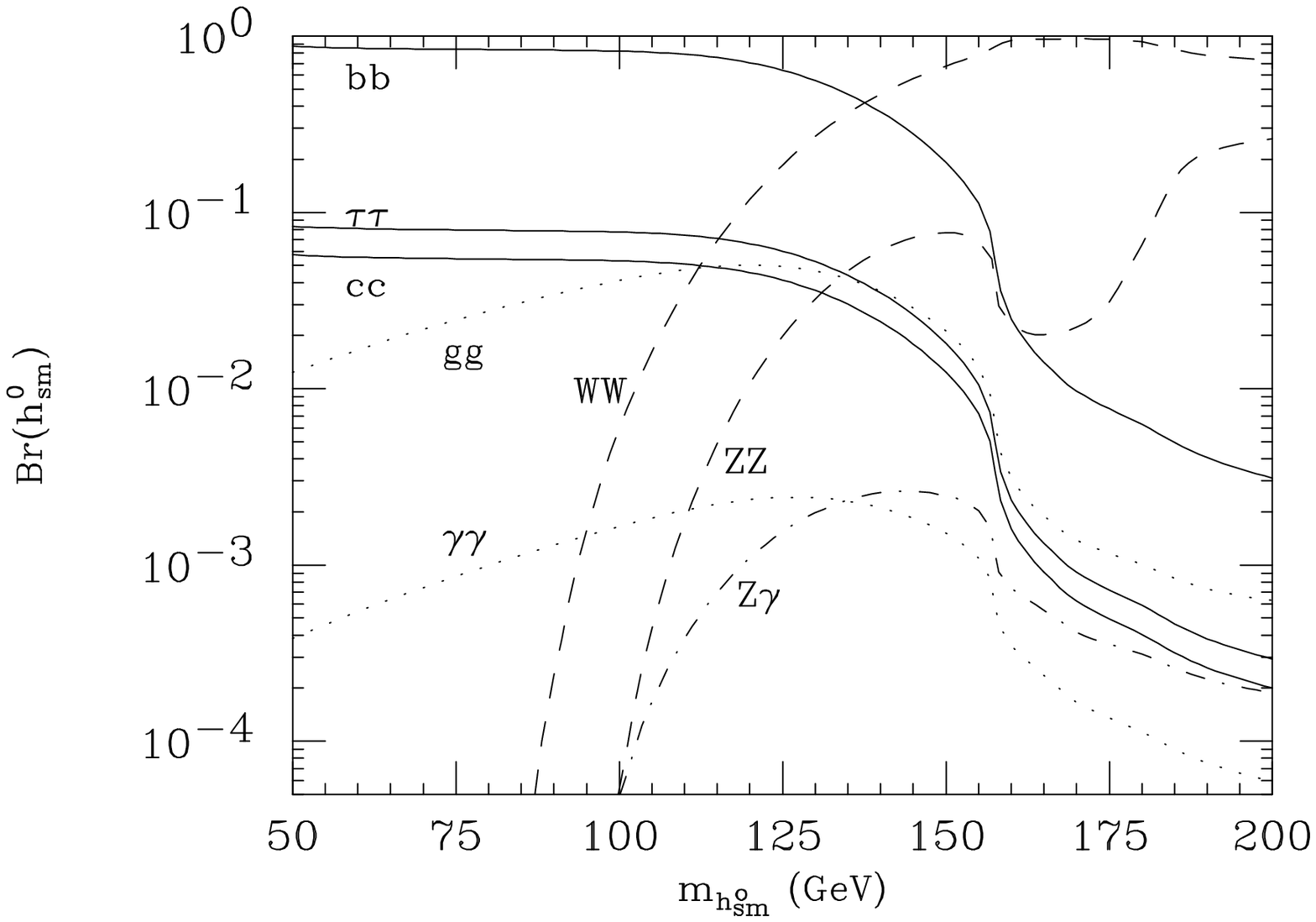}{Decay branching fractions of the standard
model Higgs boson.}

Each of the standard model decay modes plays an important role in
standard model Higgs phenomenology.  For example, the $\hsmz\to gg$
decay is directly related to the $gg\to \hsmz$ production cross-section
which is so important for the LHC searches.   Likewise the 
$\hsmz\to \gamma\gamma$ mode is important as a decay mode since
a peak of the $\gamma\gamma$ distribution can be resolved from
background with good photon resolution.
At leading order 
\beq
\sigma (gg\to \hsmz\to \gamma\gamma)\propto \Gamma (\hsmz \to gg)
\Gamma (\hsmz\to \gamma\gamma).
\eeq

The branching fraction into $ZZ^{(*)}$ becomes important for the
``gold-plated'' $4l$ Higgs decay mode with the invariant mass of
all four leptons reconstructing the Higgs mass.  At the Tevatron
and also at the LHC the $W^\pm \hsmz\to l\nu b\bar b$ mode is
relevant because of the large $\hsmz \to b\bar b$ branching fraction
for a Higgs with mass below $130\gev$.  

The $gg$ and $\gamma\gamma$ branching fractions are loop mediated
decays, and it is quite likely that they could be substantially modified
if there were new physics running around the loops.  In the case of 
supersymmetry, the present mass limits preclude the possibility of 
substantial alterations of standard model predictions from new particles
in the loops.  The most important effect in the MSSM is the extended
Higgs sector which allows more complicated couplings to the standard
model particles through mixing of the additional Higgses.  This is especially
true for large $\tan\beta$ models.  Although large $\tan\beta$ models
are somewhat finetuned, they nevertheless are required by some
attractive $SO(10)$ grand unified models with $b-\tau -t$ Yukawa
unification~\cite{anderson}.

The neutral scalar Higgs mass matrix is
\beq
M^2= \left( \begin{array}{cc}
  m_A^2\sin^2\beta+m^2_Z\cos^2\beta & -\sin\beta\cos\beta (m_A^2+m_Z^2) \\
 -\sin\beta\cos\beta (m^2_A+m_Z^2) & m_A^2\cos^2\beta +m^2_Z\sin^2\beta
 \end{array} \right) +
 \left( \begin{array}{cc}
  \Delta_{11} & \Delta_{12} \\
  \Delta_{12} & \Delta_{22} 
 \end{array} \right) .
\eeq
The first term is the tree-level contribution to the mass matrix and
the second term is the one loop contribution to the mass matrix.  The
values of the $\Delta$'s can be found in several places, 
including in~\cite{ellis}.  
As $\tan\beta$ gets higher the value of $m^2_A\sin\beta\cos\beta$
gets smaller, and typically the value of $\Delta_{12}$ gets larger.
Thus, it is possible to have a cancellation between the tree-level contribution
and the one-loop correction such that $M^2_{12}=0$. 
In this extreme case, the
two eigenvalues are pure $h^0_u$ and $h^0_d$ with masses
\bea
m^2_{h^0_d} & = & -m^2_Z|\cos 2\beta | +\Delta_{12}\tan\beta +\Delta_{11} \\
m^2_{h^0_u} & = & m^2_Z|\cos 2\beta | +\frac{\Delta_{12}}{\tan\beta}
                        +\Delta_{22}.
\eea
This limit is only possible for $\tan\beta \gsim 30$, and so the
lightest eigenvalue is well approximated by the value
$m_Z^2+\Delta_{22}$ which implies $m_{h^0_u}\gsim 100\gev$ given
current squark bounds.  
Furthermore $\langle H_u^0\rangle =v_u\simeq v$, and so
$h^0_d$ acts as a spectator to electroweak symmetry
breaking.

In Fig.~\ref{br-hu} we plot the decay branching fractions for an extreme
case where $\hsmz\simeq h^0_u$, versus $m_{h^0_u}$.
Since the $h^0_u$ has no tree-level couplings to
the $b$ quarks its partial width into them is negligible.  However,
the partial widths to $gg$, $\gamma\gamma$, $WW$, etc.\ are not appreciably
affected. 
\jfig{br-hu}{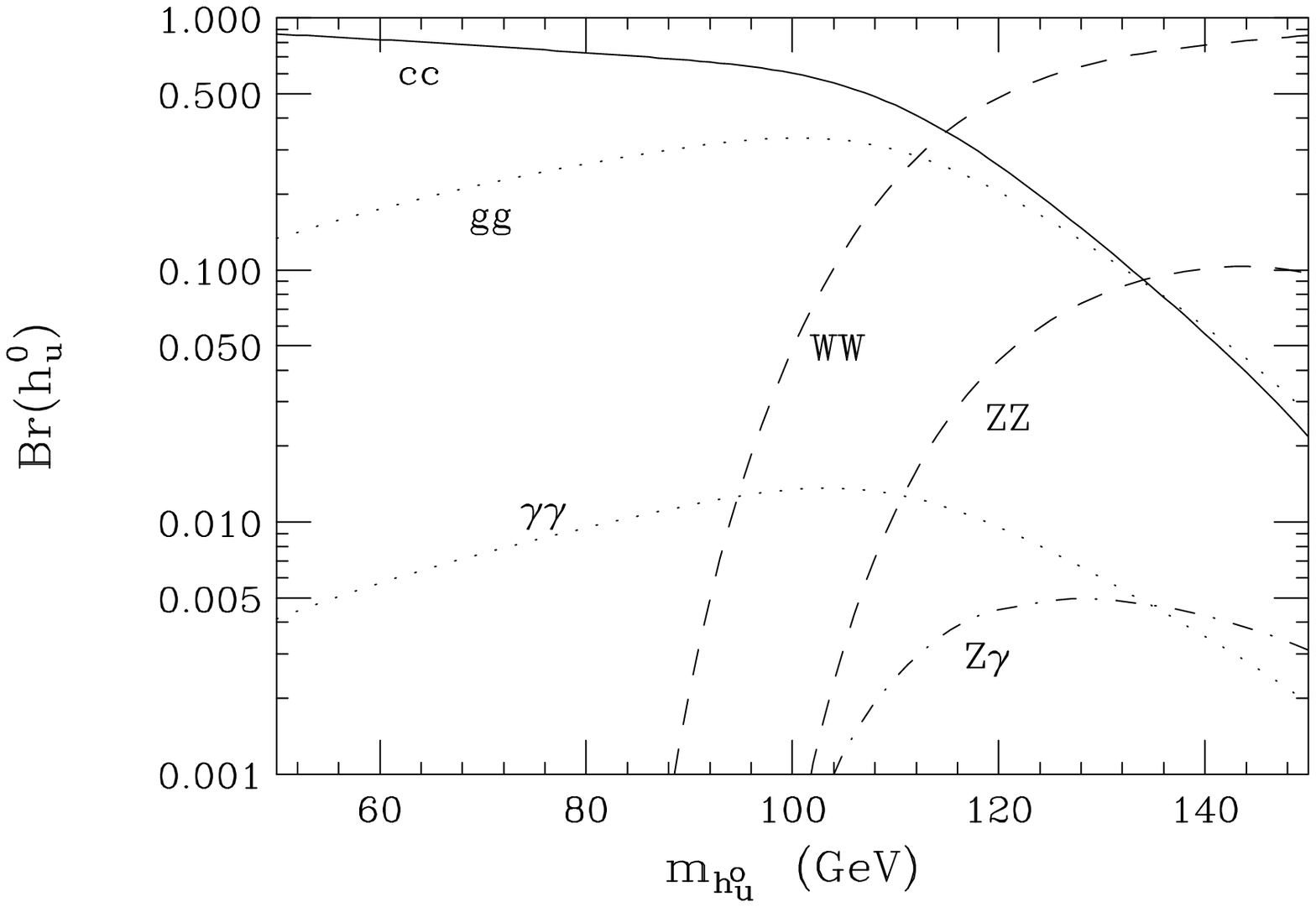}{Decay branching fractions of $h^0_u$ Higgs boson.}
Therefore, the branching fraction into $WW^*$ becomes 
even more significant.  For a Higgs mass of $100\gev$ the $h^0_u$
branching fraction into $WW^*$ is more than eight times that
of $\hsmz$. For a Higgs mass of $120\gev$ this enhancement factor drops
down to about four times, and for $140\gev$ it is a little less than
a factor of two enhancement.

This suppression of the $b\bar b$ mode 
for high $\tan\beta$ models is somewhat generic
even when the cancellation between the tree level and one loop corrections
are not exact~\cite{kane}.  
This is true when the tree-level and one-loop contributions
are of opposite sign to each other, usually implying a particular
sign for the $\mu$ parameter.  The other sign of the $\mu$ parameter will
generally lead to a much enhanced $b\bar b$ Higgs branching mode in
supersymmetric models, making the $WW^*$ decay mode less important.
Discussion of supersymmetry and the light Higgs together is appropriate
since it can be shown that even the most complicated Higgs sector
will yield at least one light eigenvalue below about $150\gev$ provided
the gauge couplings remain perturbative up to the GUT/Planck 
scale~\cite{kane93}.

Many models which have
more than one degree of freedom
contributing to electroweak symmetry breaking 
will yield non-standard model like branching fractions for the lightest
Higgs particle, and can yield an enhanced $WW^*$ branching fraction.  
The effect is most pronounced if fermion mass
generation is largely separated from electroweak symmetry breaking --
a possibility that is certainly not unreasonable~\cite{wells}.  
In this case at least
one scalar will have larger branching fractions into vector bosons
than the standard model Higgs.  In the extreme case of a ``bosonic''
Higgs, the branching fractions to $\gamma\gamma$ dominate with
Higgs mass below $90\gev$ and the $WW^{(*)}$ branching fractions
begin to dominate for Higgs mass above that~\cite{bosonic,stange}.
Many of the results that
we will show below will be using the standard model as a primary
example, but it should always be kept in mind that
the significance of the trilepton Higgs signal can
be greatly enhanced if nature chooses a somewhat more complicated
symmetry breaking pattern.

\section{Trilepton Higgs signal at the Tevatron}
\label{tevatron}
\bigskip

Since single production of $\hsmz$ via $gg\rightarrow \hsmz$ does not lead
to observable signals at the Fermilab Tevatron collider, the 
main production mode of interest for the Higgs boson is 
$p\bar p \to W^*\to W^\pm \hsmz$\cite{stange}. 
In the introduction we noted that
the $l\nu b\bar b$ decay mode of this process has been studied in
detail and it has been determined that with $30\xfb^{-1}$ of data one could
detect the Higgs boson if it had mass below $120\gev$.  Above
$120\gev$ not only does the cross-section decrease but the branching
fraction into $b$ quarks drops as well.  If there is any hope to see
the standard model Higgs at the higher masses, one will need to
study the more dominant decay mode $\hsmz\to WW^*$ and hope that
a signal above background could be found.

The subprocess total cross section for $W\hsmz$ is given by
\beq
\hat{\sigma}(d\bar{u}\rightarrow W^-\hsmz )=
\frac{g^4 M_W^2\lambda^{1\over 2}(\hat{s},m_{\hsmz}^2,M_W^2) |D_W(\hat{s})|^2}
{192 \pi\hat{s}}\left( 1+
\frac{\lambda (\hat{s},m_{\hsmz}^2,M_W^2)}{12\hat{s}M_W^2}\right),
\eeq
where $\lambda (a,b,c)=a^2+b^2+c^2-2ab-2ac-2bc$ and 
$D_W(\hat{s})=1/(\hat{s}-M_W^2+iM_W\Gamma_W)$.
We convolute the above subprocess cross section with CTEQ4 parton 
distribution functions, and scale our results so that they are
in accord with
recently calculated NLO QCD cross section calculations\cite{qcd}.
In Fig.~\ref{tev} the upper line 
shows the total next-to-leading order cross section
for $p\bar p\to W^\pm \hsmz$ at the Tevatron with $\sqrt{s}=2\tev$.
Below that we have multiplied the cross-section by the branching
fraction of the Higgs to decay to $WW^{(*)}$ and also multiplied by
the probability that each of the three $W$'s in the event will decay
to $e\nu$ or $\mu\nu$, thus yielding a trilepton signal.
\jfig{tev}{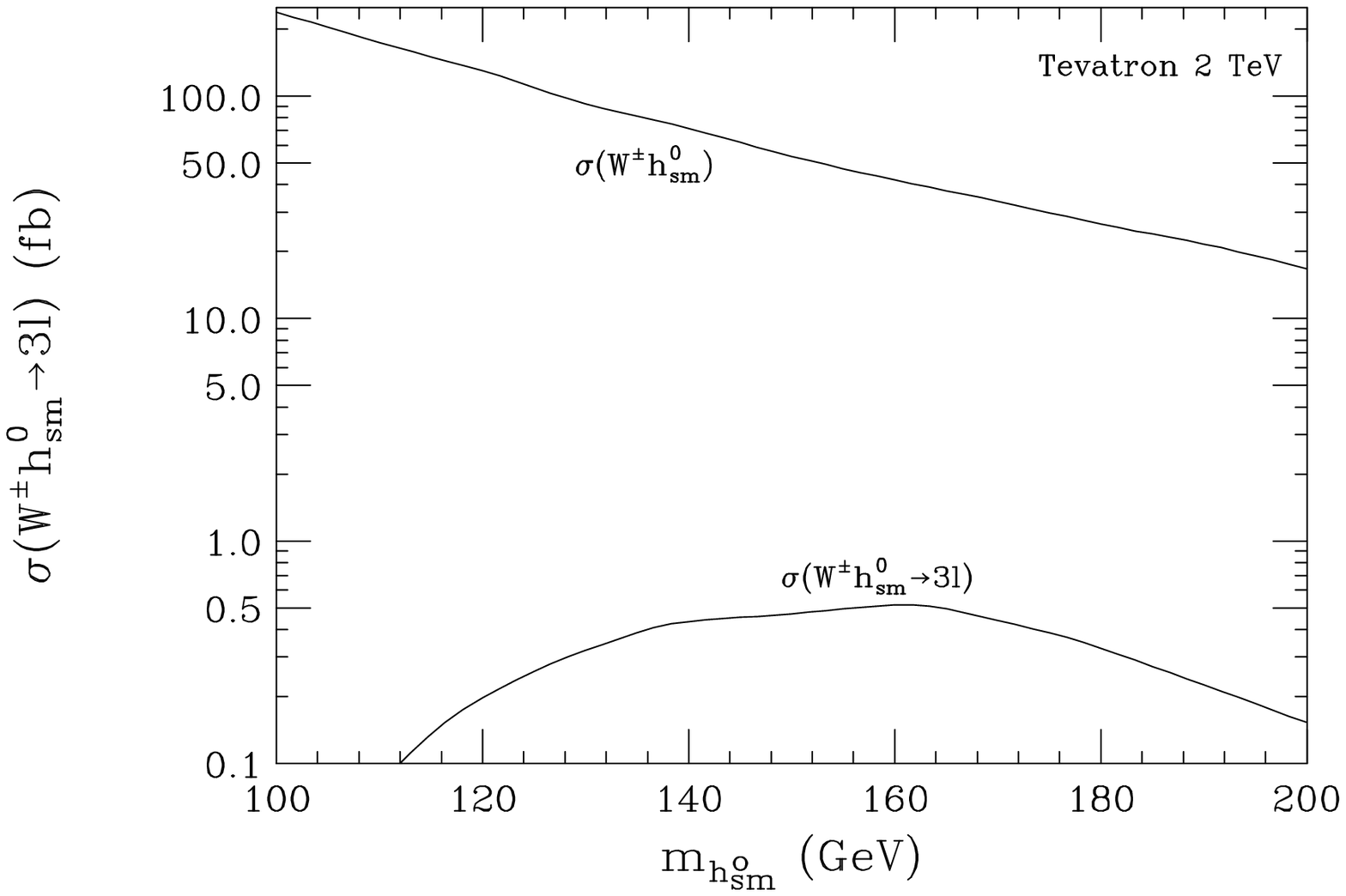}{The upper line is the total next to leading order
$W^\pm \hsmz$ cross section at the Tevatron with $2\tev$ center of mass
energy.  The lower line is the total three lepton cross section
from $W^\pm \hsmz\to 3l$.}

To calculate differential distributions, we calculate
$q\bar{Q}\rightarrow l\bar{\nu}\hsmz$ production using Monte Carlo
integration.
Spin correlation effects are included by using the squared matrix 
elements given by
Baer {\it et al.}\cite{qcd}. For $\hsmz\rightarrow WW\rightarrow l_1\bar{\nu_1}
\bar{l_2}\nu_2$ decay, we implement the spin correlated matrix element given
by Barger {\it et al.}\cite{barger90}.
The three final state leptons are generally highly energetic and isolated.
The events also have a substantial amount of missing $E_T$ associated
with them.  To demonstrate this, we show in Fig.~\ref{pttev}
the lepton $p_T$ and missing $E_T$ for $m_{\hsmz}=160\gev$.
\jfig{pttev}{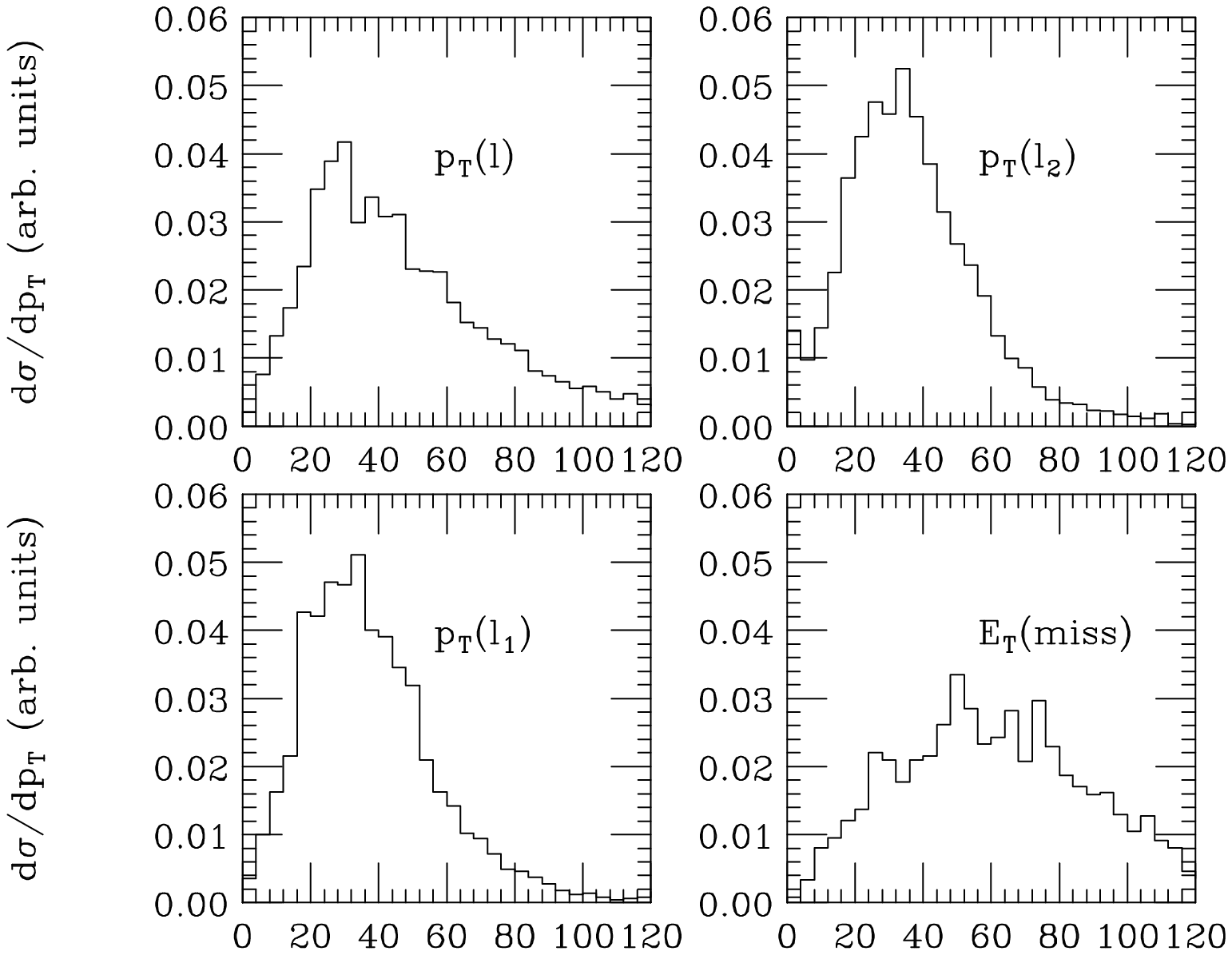}{Lepton $p_T$ and missing $E_T$ distributions
from $W^\pm \hsmz\to 3l$ at the Tevatron.  The Higgs mass is $160\gev$ in
this figure.}

At the Tevatron the main background is
expected to be from $WZ$ production~\cite{baer95} 
where the $W$ decays to $e$ or $\mu$, and
the $Z$ decays to two leptons. We include QCD corrections to our $WZ$
background estimate\cite{ohnemus}.
To reduce backgrounds, we employ
the following cuts~\cite{baer95}: (1) Three isolated 
and central ($|\eta |<2.5$) leptons with
$p_T=(20, 15, 10)\gev$ for the three leptons in descending order of $p_T$;
(2) $\slashchar{E_T}>25\gev$; (3) the invariant mass of the opposite-sign,
same-flavor lepton pair not reconstruct to the $Z$ boson mass,
$|m_{l^+l^-}-m_Z|< 10\gev$; (4) no jets in the event. The jet veto 
typically reduces purely electroweak sources of trileptons by a 
factor of 2\cite{baer95}, while reducing $t\bar{t}$ background to levels
well below that from $WZ$ production. 
We incorporate a jet veto factor of 0.5 in our parton level signal generator. 
After all the cuts are applied the background is reduced to $0.25\xfb$.

In Fig.~\ref{sigtev} we plot the trilepton signal after cuts.  
\jfig{sigtev}{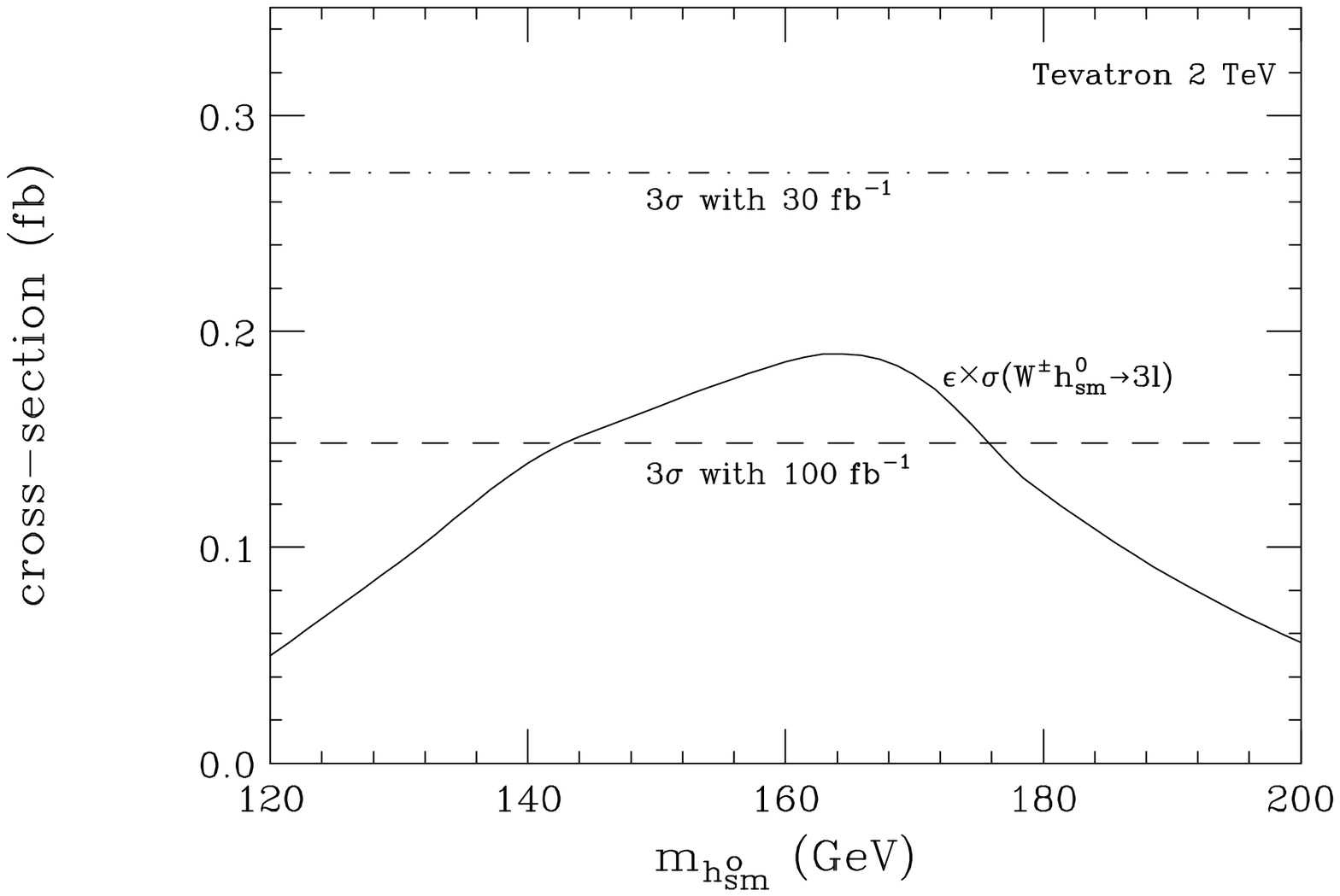}{The solid line is the total Higgs trilepton
cross-section after background reducing cuts.  The dash-dotted line
is the $3\sigma$ discovery contour given $30\xfb^{-1}$ of data, and
the lower line is the $3\sigma$ discovery contour given $100\xfb^{-1}$
of data.}
The reduction of the signal varies
from $0.25$ to $0.41$ in the plotted Higgs mass range.  Also in
Fig.~\ref{sigtev} we have put $3\sigma$ discovery contours for
$30\xfb^{-1}$ and $100\xfb^{-1}$.  With $30\xfb^{-1}$ a $3\sigma$ confidence
on a discovery does not appear possible.  With $100\xfb^{-1}$ the
$3\sigma$ range corresponds to about $140\gev\lsim m_{h^0_{sm}}\lsim 175\gev$.

The trilepton Higgs signal significance is plotted versus $m_{h^0}$ for
different Higgs models in Fig.~\ref{sigma-tev}.
\jfig{sigma-tev}{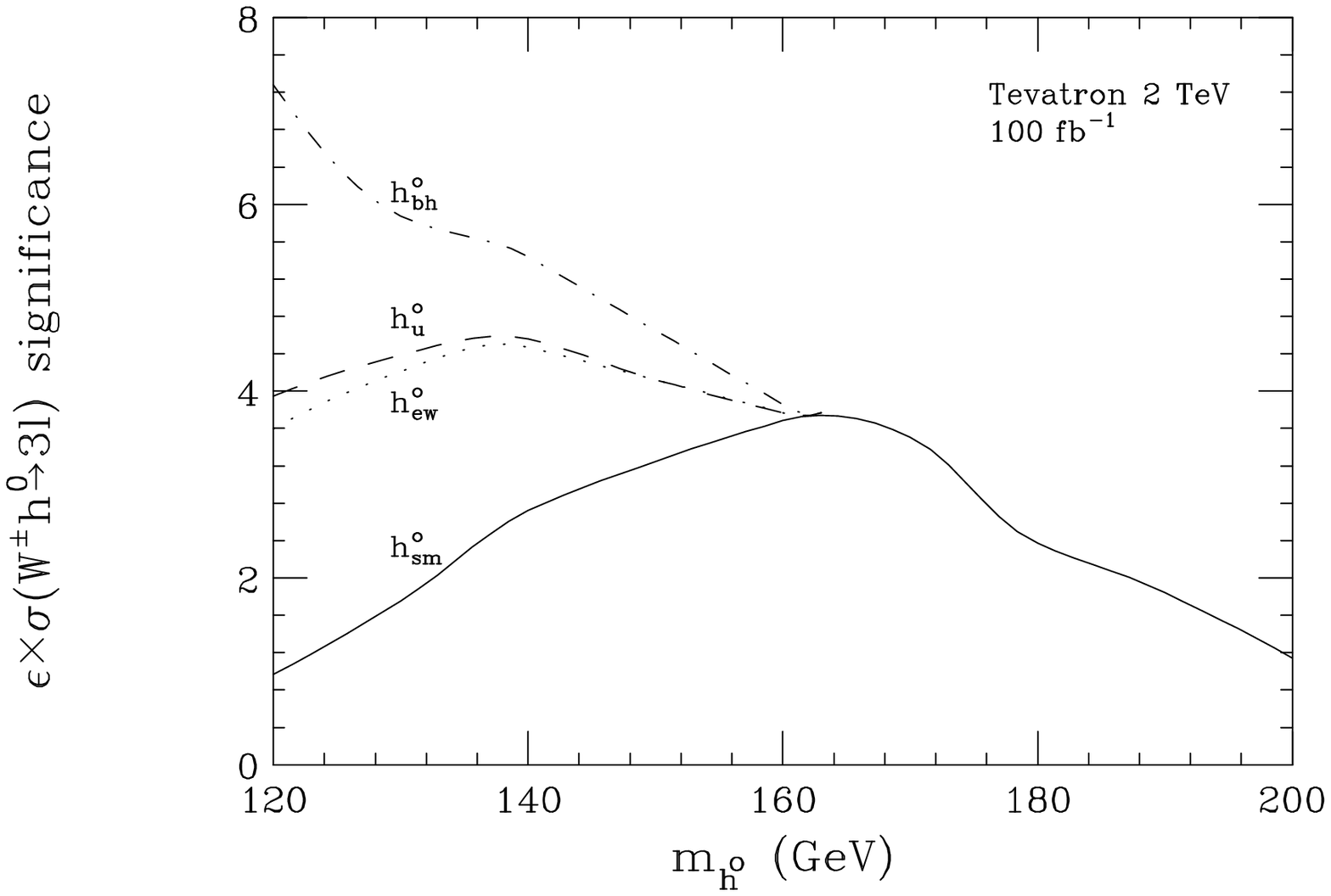}{The discovery significance for different
Higgs models as a function of the Higgs mass.  $h_{bh}^0$ is the purely
bosonic coupled Higgs boson; $h_{ew}^0$ is the electroweak symmetry breaking
Higgs boson of a top quark condensate model; $h^0_u$ is the ``up-Higgs'' of
high $\tan\beta$ supersymmetry models which do not couple to down quarks
or leptons; and, $h^0_{sm}$ is the standard model Higgs boson.}
$h_{bh}^0$ is the purely
bosonic coupled Higgs boson~\cite{bosonic,stange}; 
$h_{ew}^0$ is the electroweak symmetry breaking
Higgs boson of a top quark condensate model~\cite{wells}; 
$h^0_u$ is the ``up-Higgs'' of
high $\tan\beta$ supersymmetry models which do not couple to down quarks
or leptons; and, $h^0_{sm}$ is the standard model Higgs boson.  The
non-standard Higgses have higher significance at lower Higgs mass because
the branching fraction $B(h^0\to WW^*)$ is much higher for these Higgses
than the standard model Higgs.  When $160\gev \lsim m_{h^0}\lsim 200\gev$ 
the branching 
fractions into $WW^*$ are about the same for all Higgs models.

Even when the signal significance is marginal, it still might be possible
to extract evidence for a Higgs boson by looking 
carefully at the lepton kinematics on
an event by event basis.  For example, the main background arises from
$WZ$ when the $Z$ decays to $\tau^+\tau^-$ and the $\tau$'s subsequently
decay leptonically.  In this case, the leptons from the $\tau$ decay will
be softer than any of the leptons in the signal.  One variable that might
be useful to analyze is what we call $p_T(OS)$.  It is defined to be
\beq
p_T(OS)=\min \{ p_T(l^\pm)+p_T(l_1^\mp), p_T(l^\pm)+p_T(l_2^\mp) \}.
\eeq
Among the background $WZ\to l^\pm l_1^\mp l_2^\mp$ event sample, the
unique charge lepton ($l^\pm$) must come from a lepton in the $Z$
decay.  The other two leptons with opposite charge come either from the
$W$ or $Z$.  Generally, the lepton from the $Z$ will be softer than the one 
from the $W$.  Therefore, $p_T(OS)$ usually sums the $p_T$'s of the
two leptons which come from $Z\to \tau^+\tau^-\to l^+l^-$.
In Fig.~\ref{ptos} we plot the distribution of $p_T(OS)$ for the
$WZ$ background (dashed line) and the signal (solid line)
with $m_{\hsmz}=160\gev$. 
\jfig{ptos}{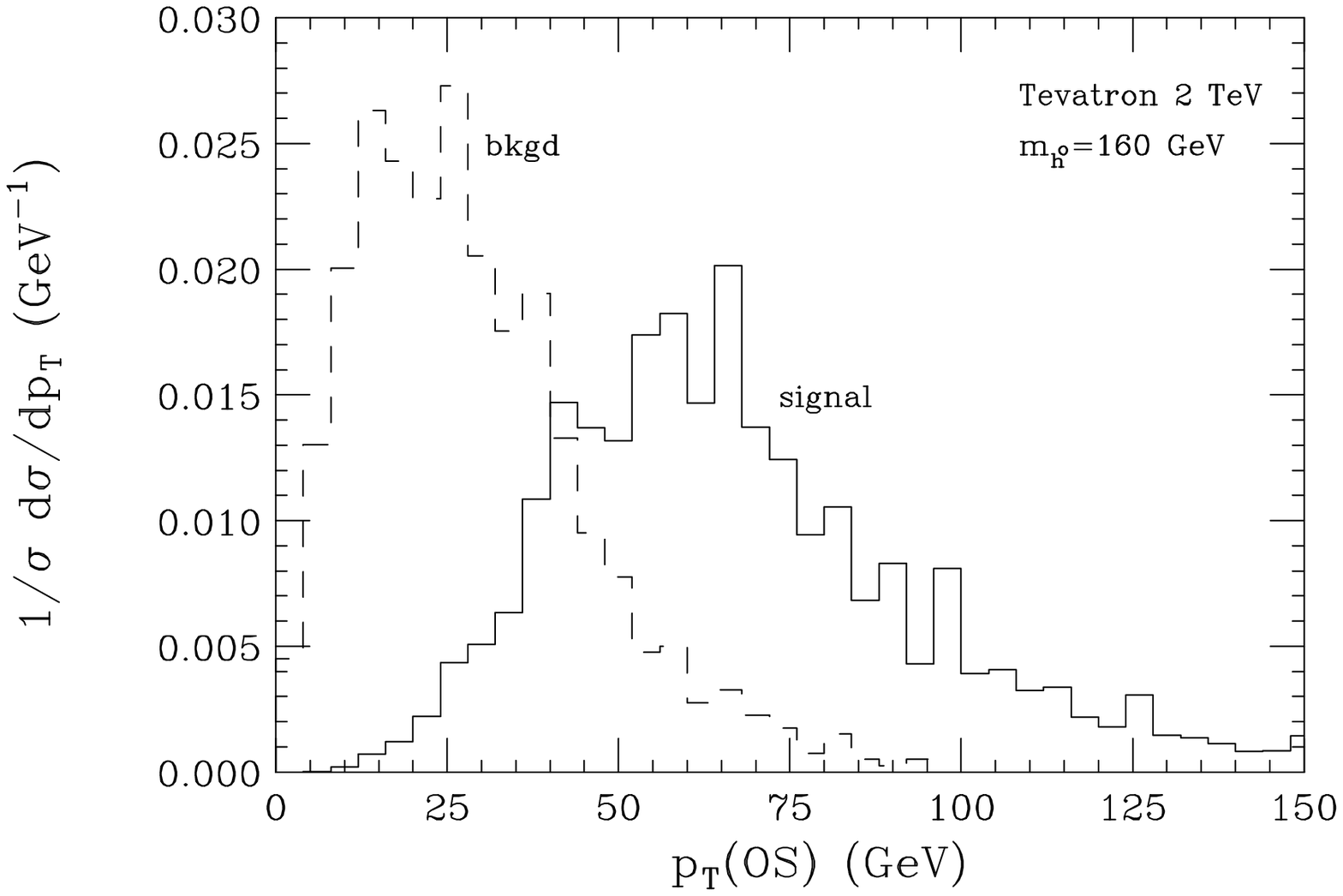}{Distribution of $p_T(OS)$, which is defined to
be the minimum
$p_T$ sum of opposite sign leptons in the $3l$ events.
The solid line is the distribution from
the $W\hsmz\to 3l$ signal, and the dashed line is the distribution
from the $WZ$ background.  Events with high $p_T(OS)$ would provide
strong hints for the $Wh\to 3l$ signal.}

{}From Fig.~\ref{ptos} it is clear that the $p_T(OS)$ spectrum for the
signal is much harder than that of the background.  The total number
of signal trilepton events expected at the Tevatron is quite small,
and as shown earlier one does not expect to get statistically compelling
signal for the standard model Higgs with less than $100\xfb^{-1}$.  However,
if there are a handful of trilepton events and several of these events
have $p_T(OS)\gsim 70\gev$ then this would be a powerful hint for the
presence of the signal.  If several events showed up with $p_T(OS)>100\gev$
then the possibility that it came from $WZ$ background fluctuations
would be extremely small, and the evidence for a signal would be
intriguing.  To get two or more signal
events with such large $p_T(OS)$ would also need to be 
somewhat of a lucky fluctuation.
With $30\xfb^{-1}$ the expected number of signal events after all cuts
and with $p_T(OS)>100\gev$ is about $0.8$ events for $m_{\hsmz}=160\gev$.  
Therefore, a fluctuation of two or more signal events with $p_T(OS)>100\gev$ 
has a reasonable probability of occurring, whereas the background has a 
negligible probability of producing two or more such events.

\section{Trilepton Higgs signal at the LHC}
\bigskip

At the LHC the $gg\to \hsmz\to \gamma\gamma$ mode is perhaps the most 
promising way to discover an intermediate mass SM Higgs boson.
Indeed, even the lightest supersymmetric Higgs has excellent prospects
for being discovered in this mode~\cite{baer91,kane}.  
However, other modes have been
studied, and it has been shown, for example, that 
$W^\pm\hsmz\to l\nu b\bar b$
may also be utilized to detect a Higgs if its mass is below about
$120\gev$. 

The trilepton Higgs signal at the LHC provides a complementary way 
to detect a Higgs boson in the intermediate mass region. 
In Fig.~\ref{lhc} the upper line
plots the total next-to-leading order $W^\pm \hsmz$ production cross-section
for the standard model Higgs at the LHC with $\sqrt{s}=14\tev$.  
\jfig{lhc}{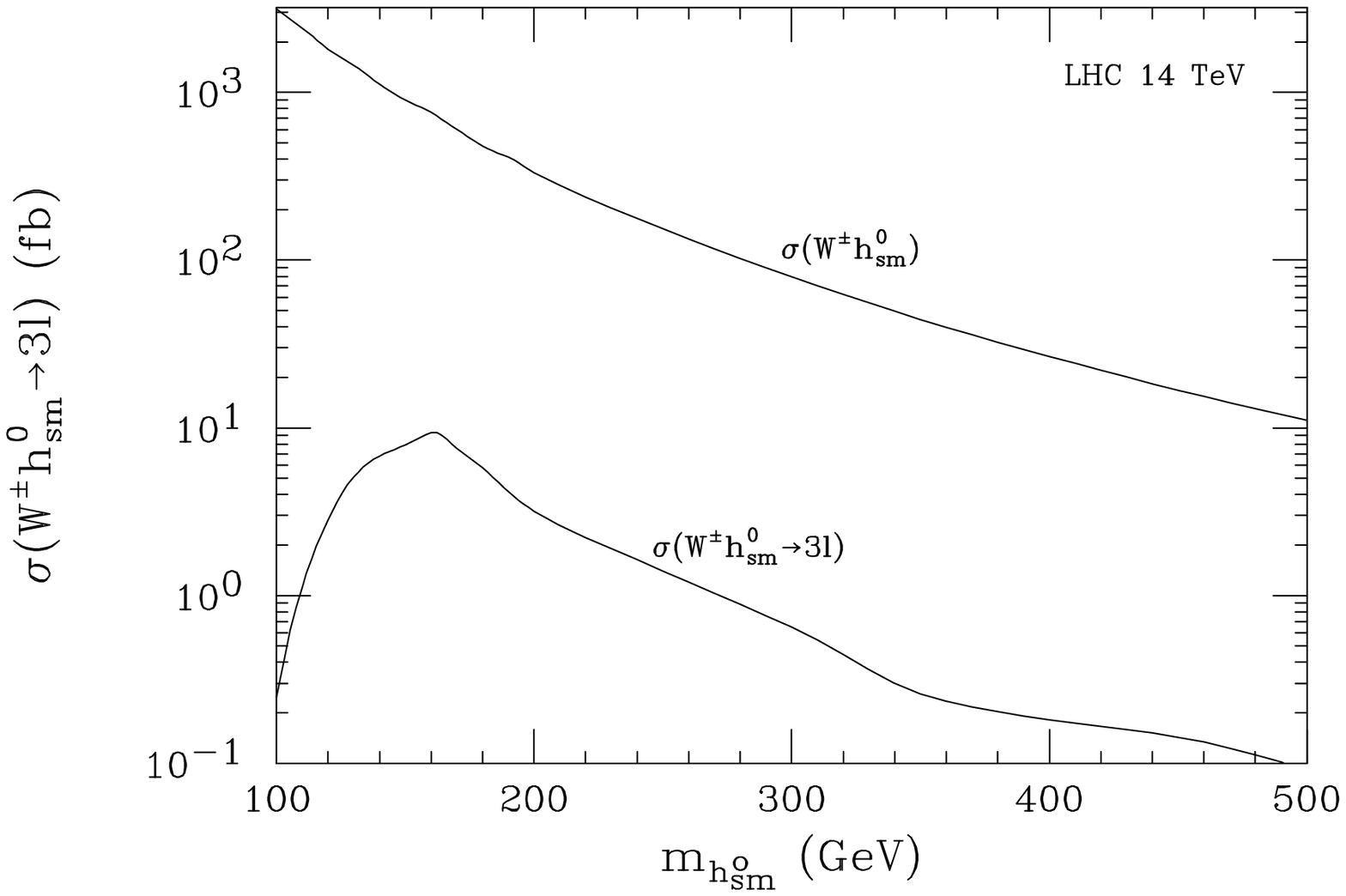}{The upper line is the total next to leading order
$W^\pm \hsmz$ cross section at the LHC with $14\tev$ center of mass
energy.  The lower line is the total three lepton cross section
from $W^\pm \hsmz\to 3l$.}
The lower line is the total three lepton cross-section after all branching
fractions have been folded in. In Fig.~\ref{ptlhc} we show the lepton
$p_T$ and missing $E_T$ distributions for $m_{\hsmz}=160\gev$.
\jfig{ptlhc}{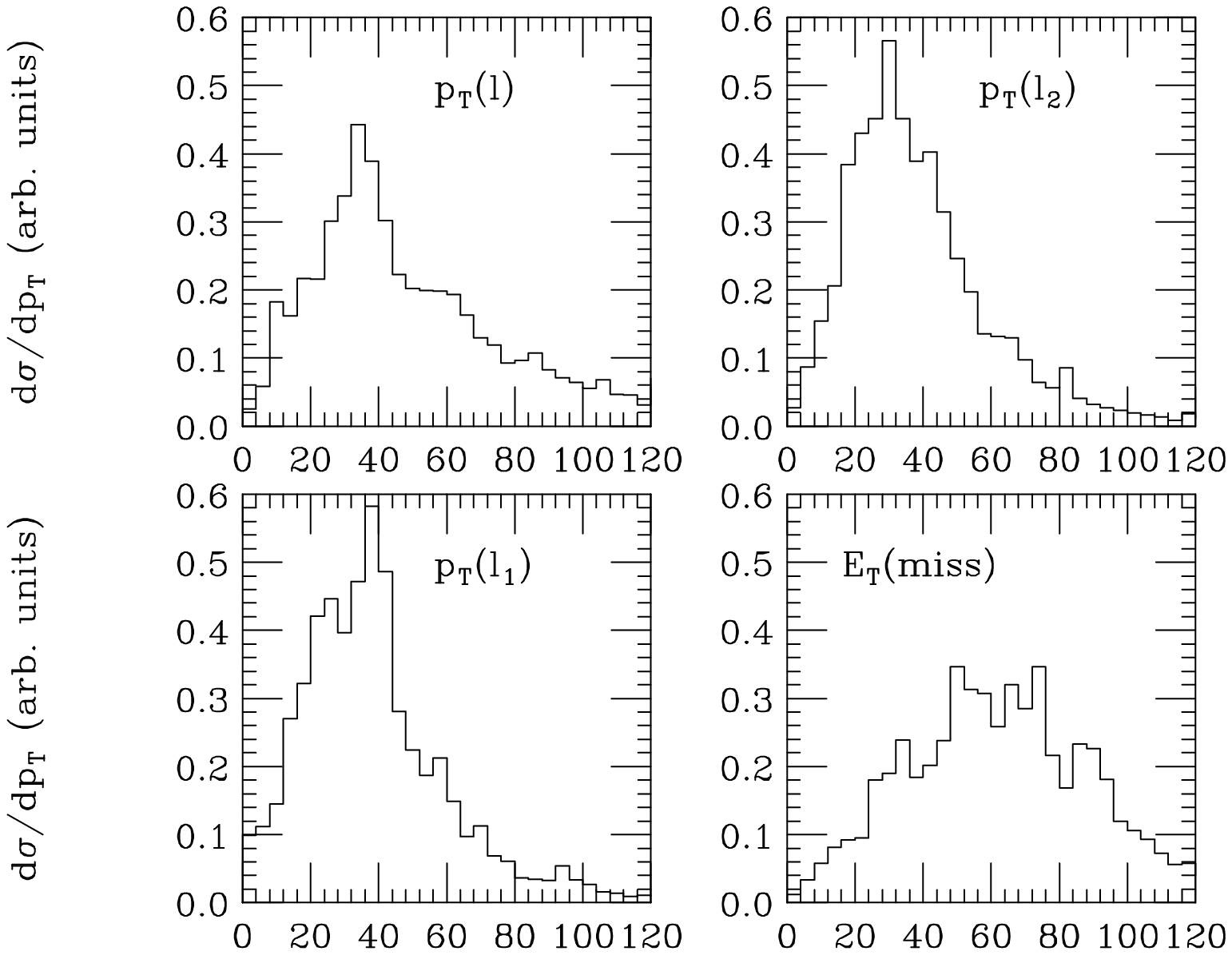}{Lepton $p_T$ and missing $E_T$ distributions
from $W^\pm \hsmz\to 3l$ at the LHC.  The Higgs mass is $160\gev$ in
this figure.}

The lack of hadronic activity in
the partonic subprocess for the signal provides a useful tool for
overcoming backgrounds.  The main background at the LHC is
again $t\bar t$ and $WZ$ production.  These backgrounds
can be reduced by requiring\cite{baer94} 
(1) $p_T(l_1,l_2,l_3)>(20, 20, 10)\gev$,
(2) no central ($|\eta |<3$) jets with $p_T>25\gev$, (3) 
$|m_{l^+l^-}-m_Z|< 8\gev$.  These constitute the ``soft'' cuts against
background of Ref. \cite{baer94}, 
leaving a total background rate of $4.3\xfb$.  
The ``hard'' cuts which further reduce the $t\bar t$ 
background keep events only if the two fastest leptons are the
same sign and the third lepton has the same flavor as either
the first or second lepton, {\it or} if the two fastest leptons have
opposite sign and the third lepton has $p_T>20\gev$.  With these
``hard'' cuts the background is reduced to $1.1\xfb$ which mainly comes 
from $WZ$ production where 
$Z\rightarrow\tau\bar{\tau}\rightarrow l\bar{l'}+\nu's$.

In Fig.~\ref{sigLHC-hard} we have plotted the
next-to-leading order
signal cross section after employing the cuts described in the
previous paragraph.
\jfig{sigLHC-hard}{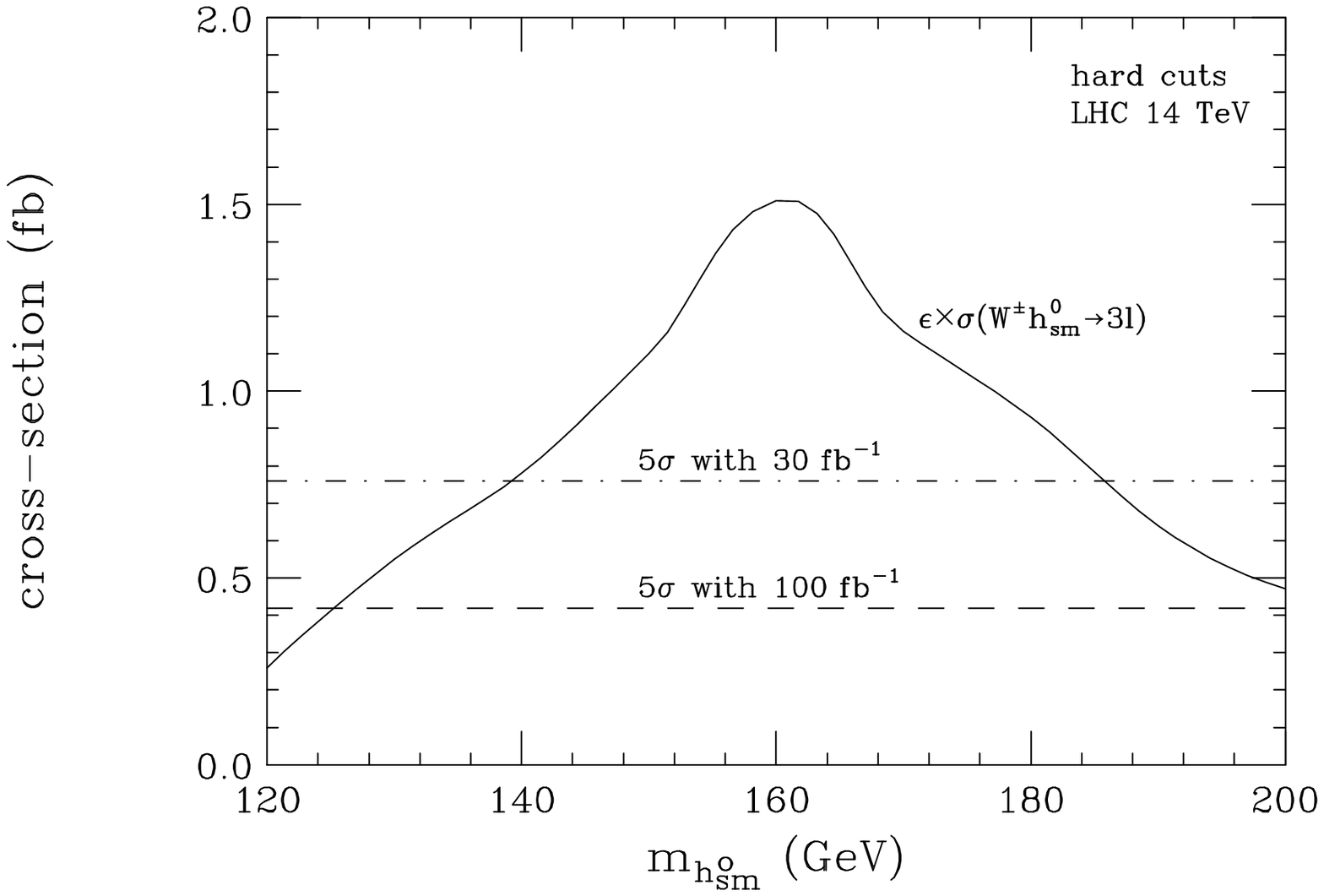}{Trilepton Higgs cross-section
after ``hard'' background reducing cuts.  The $5\sigma$ discovery contours
are shown assuming $30\xfb^{-1}$ and $100\xfb^{-1}$ of integrated
luminosity.}
The $5\sigma$ discovery
contours are shown assuming $30\xfb^{-1}$ and $100\xfb^{-1}$.  
For $30\xfb^{-1}$ ($100\xfb^{-1}$), it appears possible to see a 
Higgs trilepton signal for
$140\gev < m_{\hsmz}< 185\gev$ ($125\gev < m_{\hsmz}< 200\gev$) 
at the LHC! In addition, the Higgs trilepton signal covers the difficult
region $150\gev < m_{\hsmz}< 180\gev$ where the $\hsmz\rightarrow\gamma\gamma$
signal is rapidly diminishing, and $\hsmz\rightarrow ZZ^*$ is proceeding at a 
low rate with one off-shell $Z$-boson.

One difficulty with the trilepton Higgs signal at the LHC is that
a Higgs mass reconstruction appears difficult since there are two neutrinos
in the Higgs boson decay products. However, the invariant mass of the opposite
sign/same flavor dileptons from $\hsmz$ decay will be kinematically bounded
by $m_{\hsmz}$, and the distribution should scale with $m_{\hsmz}$.
We show the idealized distribution in Fig. \ref{mll} for several choices of 
Higgs boson mass. Fig. \ref{mll} always assumes the correct choice of 
of dilepton pair in reconstructing the invariant mass. Even for this idealized 
case, a mass reconstruction would be difficult given the expected number
of events from a signal channel.
\jfig{mll}{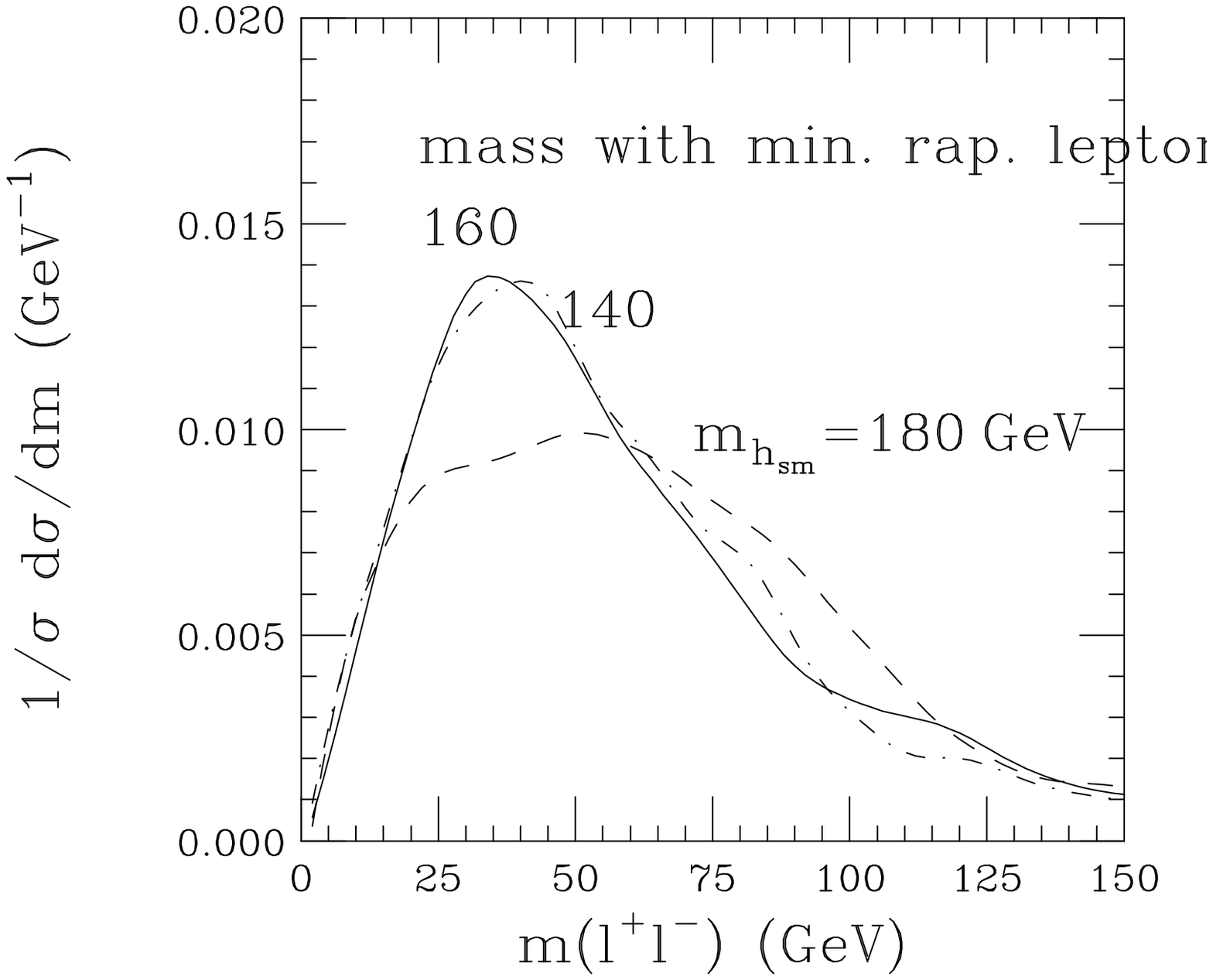}{Invariant mass of opposite-sign/same flavor
dilepton pairs from the higgs trilepton signal at the LHC for
$m_{\hsmz}=140$, 160 and 180 GeV.}

\section{Conclusion}
\bigskip

We have shown that a $3\sigma$ trilepton
SM Higgs signal can be found at the Tevatron
with $100\xfb^{-1}$ in the range $140\gev < m_{\hsmz} < 175\gev$.
Models of electroweak symmetry breaking beyond the single Higgs
postulate of the standard model may enable discovery of the Higgs
with more significance at lighter masses.  Furthermore, analyzing
the kinematics of each detected event in the trilepton sample 
would add more discriminating power to the naive significance 
values we calculated above.  Perhaps the most useful observable 
to analyze for this purpose is the $p_T(OS)$ variable
discussed in section~\ref{tevatron}.  The distribution
of $p_T(OS)$ 
is softer for the background than for the signal, and can be used
to further determine if a trilepton Higgs signal is present
in the data.

We also found that at the LHC with $30\xfb^{-1}$ ($100\xfb^{-1}$),
a $5\sigma$ discovery can be made
for the Higgs in the trilepton mode in the mass range
$140\gev\lsim m_{\hsmz}\lsim 180\gev$ \hfill\break
($125 \gev \lsim m_{\hsmz} \lsim 200 \gev$).  
Reconstructing the Higgs
mass will be difficult, although some guidance can be obtained by
examining the invariant mass distribution of opposite-sign/same flavor
dilepton pairs. Other possible techniques have been presented in
Ref. ~\cite{dittmar}. 

Other modes associated with $W^\pm \hsmz\to WWW^*$ might also be
useful to study as confirming evidence for a signal.  For example,
one might be able to see an excess in 
like-sign dilepton samples, such as $l^\pm l^\pm jj+\slashchar{E_T}$.
Furthermore, the $Z\hsmz$ production and decay could also be useful
to analyze in several channels including $Zl^+ l'^- +\slashchar{E_T}$. 
The statistical significance of these other 
modes is not as impressive as
the $3l$ signal discussed above, but it might be possible to use
them to bolster the claims for a signal in the $3l$ channel and to
get a better handle on the Higgs mass.

\section*{Acknowledgments}

We would like to thank the Aspen Center for Physics where this
work was originated.




\begin{thebibliography}{20}
\bibitem{ohnemus}
Next to leading order normalization for $W^\pm Z$ 
background at the Tevatron and LHC are extracted from
J.~Ohnemus, \PRD{44}{91}{3477}.
\end{thebibliography}
\end{document}